\documentclass[12pt]{article}
\usepackage{pdproc} 

\begin{document}
\begin{flushright}
UH511-1027-03 \\
May 2003
\\
\end{flushright}
\renewcommand{\thefootnote}{\alph{footnote}}
  
\title{ NEUTRINO DECAYS AND NEUTRINO TELESCOPES}

\author{ SANDIP PAKVASA}

\address{ Department of Physics, University of Hawaii,
  \\
 Honolulu, HI  96822, USA\\
 {\rm E-mail: pakvasa@phys.hawaii.edu}}




\abstract{It is shown how deviations of neutrino flavor mix from the
canonical 1 : 1 : 1 in high energy
astrophysical neutrinos can provide evidence for neutrino decays with lifetimes as long as 
$10^4 s/eV$.}
\normalsize\baselineskip=15pt

\section{Introduction}
 In my talk at the last Venice meeting\cite{pakvasa}, I discussed neutrino fluxes, and especially flavor composition of the neutrino beams, from astrophysical sources of high energy neutrinos.  It is generally agreed that pion and muon decays produced in beam dump type situations are the most likely sources of neutrinos in these environments\cite{learned}.  The fraction of heavy flavors, which can produce $\nu_\tau 's$, even at these high energies, is very 
small\cite{learned1}.  In this case the initial flavor mix is given to a good approximation by $\phi_{\nu_{e}} : \phi_{\nu_{\mu}} : 
\phi_{\nu_{\tau}}  \approx 1:2:0.$

With the current knowledge of $\delta m^{2}_{ij}$, and for the large L/E that we anticipate, the oscillations average out so that on arrival at earth, the flux of each mass eigenstate are given by $\phi_{\nu_{i}} = 
\sum_{\alpha} \phi^{source}_{\nu_{\alpha}} \mid U_{\alpha i} \mid^2$, 
where $U_{\alpha i}$ are elements of the neutrino mixing matrix.  For three
active neutrino species (as we assume throughout) there is now strong
evidence to suggest that $\nu_\mu$ and $\nu_\tau$ are maximally mixed and
$U_{e3} \simeq 0$.  The consequent $\nu_\mu - \nu_\tau$ symmetry means that
in the mass eigenstate basis the neutrinos are produced in the ratios
$\phi_{\nu _1}: \phi_{\nu _2}: \phi_{\nu _3} = 1:1:1$, independent of the solar mixing angle.  Oscillations do not change these proportions, but only the relative phases between mass eigenstates, which will be lost.  An incoherent mixture in the ratios 1 : 1 : 1 in the mass basis implies an equal mixture in any basis, and in particular the flavor basis in which the neutrinos are detected \cite{learned1,athar}.

There is also the possibility that the initial flavor mix is not the canonical $1:2:0$.  For example, the muons can lose energy (as at high energies in the atmosphere) e.g. in a magnetic field; in that case no high energy $\nu_e's$ are available and the flavor mix becomes \cite{rachen}
$\phi_{\nu_{e}} : \phi_{\nu_{\mu}} : \phi_{\nu_{\tau}} = 0 :1: 0$.
The effect of mixing then is to produce a final mixture which is given by $\phi_{\nu_{e}} : \phi_{\nu_{\mu}} : \phi_{\nu_{\tau}} = a : 1 : 1$ with $a$ between 1/2 to 2/3 depending on the precise value of the solar mixing angle.

This shows the value of measuring the flavor mix; it can yield information
on the production mechanism and the environment; even in the presence of
mixing and oscillations.  But it can do even more; such as yield information
on neutrino properties.  There are other ways in which the flavor mix can change from the canonical $1:1:1$ figure. One in particular, which gives rise to striking signatures, is the decay of neutrinos\cite{beacom}.  Before discussing these, let me motivate the possibility of neutrino decay. 

\section{Neutrino Decay\cite{pakvasa1}}
  We now know that neutrinos have non-zero masses and non-trivial mixings.  This is based primarily on the evidence for neutrino mixings and oscillations from the data on atmospheric neutrinos and on solar neutrinos.

If this is true then in general, the heavier neutrinos are expected to decay into the lighter ones via flavor changing processes.  The only questions are (a) whether the lifetimes are short enough to be phenomenologically interesting (or are they too long?) and (b) what are the dominant decay modes.

Throughout the following discussion, to be specific, I will assume that the
neutrino masses are at most of order of a few eV.  Since we are interested
in decay modes which are likely to have rates (or lead to lifetimes)  which are phenomenologically interesting,  we can rule out several classes of decay nodes.

First, consider radiative decays, such as $\nu_e \rightarrow \nu_j 
+ \gamma$.  Since the experimental bounds on $\mu_{\nu _i}$, the magnetic
moments of neutrinos, come from reactions such as $\nu_e e \rightarrow e
``\nu''$ which are not sensitive to the final state neutrinos; the bounds
apply to both diagonal as well as transition magnetic moments and so can be
used to limit the corresponding lifetimes. The bounds should really be on
mass eigenstates\cite{john}, but since the mixing angles are large, it does not matter
much.  The current bounds are\cite{these}:
\begin{eqnarray}
\tau_{\nu_ e}  > & 5.10^{18} \ \mbox{sec}  \nonumber \\
\tau_{\nu_\mu}   > & 5.10^{16} \ \mbox{sec} \nonumber \\
\tau_{\nu_\tau}  > & 2.10^{11} \ \mbox{sec}.
\end{eqnarray}

There is one caveat in deducing these bounds.  Namely, the form factors are
evaluated at $q^2 \sim O (eV^2)$ in the decay matrix elements whereas in the scattering from which the bounds are derived, they are evaluated at $q^2 \sim O (MeV^2)$.  Thus, some extrapolation is necessary.  It can be argued that, barring some bizzare behaviour, this is justified\cite{frere}.

An invisible decay mode with no new particles is the three body decay 
$\nu_i \rightarrow \nu_j \nu_j \bar{\nu}_j$.  Even at the full 
strength of Z coupling, this yields a lifetime of $2.10^{34}$s, far too long to be of interest.  There is an indirect bound from Z decays which is weaker but still yields 2.10$^{30}$s \cite{bilenky}.

Thus, the only decay modes which can have interestingly fast decays rates are two body modes such as $\nu_i \rightarrow \nu_j + x$ and $\nu_i \rightarrow \bar{\nu}_j + x$ where $x$ is a very light or massless particle, e.g. a Majoron.

The only possibility for fast invisible decays of neutrinos seems to lie 
with Majoron models\cite{pakvasa1}.  There are two classes of models; the I=1 Gelmini-Roncadelli\cite{gelmini} majoron and the I=0 Chikasige-Mohapatra-Peccei\cite{chicasige} majoron. In general, one can choose the majoron to be a mixture of the two; furthermore the coupling can be to flavor as well as sterile neutrinos.  The effective interaction is of the form:
\begin{equation}
g _\alpha \bar{\nu}^c_\beta \nu_\alpha \ J
\end{equation}
giving rise to decay:
\begin{equation}
\nu_\alpha \rightarrow \bar{\nu}_\beta \ ( or  \ \nu_\beta)  +  J
\end{equation}
where $J$ is a massless $J= 0 \ L =2$ particle; $\nu_\alpha$ and $\nu_\beta$
are mass eigenstates which may be mixtures of flavor and sterile neutrinos.
Models of this kind which can give rise to fast neutrino decays and satisfy the bounds below have been discussed\cite{valle}.  These models are unconstrained by $\mu$ and $\tau$ decays which do not arise due to the $\Delta L = 2$ nature of the coupling.  The I=1 coupling is constrained by the bound on the invisible
$Z$ width; and requires that the Majoron be a mixture of I=1 and
I=0\cite{choi}.
 The couplings of $\nu_\mu$ and $\nu_e$ $(g_\mu$ and $g_e)$ are constrained
by the limits on multi-body $\pi$, K decays
$\pi \rightarrow \mu \nu \nu \nu$ and $K \rightarrow \mu \nu \nu \nu$ and
on $\mu-e$ university violation in $\pi$ and K decays\cite{barger}, but not
sufficiently strongly to rule out fast decays.
Direct limits on such decay modes are also  very weak.  The current strongest bound come from solar neutrino results and are of order 
of $\tau/m \geq 10^{-4} s/eV$\cite{beacom1}.  
This limit is based primarily on the non-distortion of the Super-Kamiokande
spectrum\cite{fukuda}, and takes into account the potentially competing
distortion caused by oscillations (see also Ref.\cite{bandopadhyag}) as well as the appearance of active daughter neutrinos.  The SN 1987A 
data place no limit at all on these neutrino decay modes, since decay of the
lightest mass eigenstate is kinematically forbidden, and with the inherent
uncertanties in the neutrino fluxes, even a reasonable $\bar{\nu}_1$ flux alone can account for the data\cite{frieman}.

The strongest lifetime limit is thus too weak to eliminate the possibility of
astrophysical neutrino decay by a factor about $10^7 \times (L/100$ Mpc) 
$\times (10$ TeV/E).  Some aspects of the decay of high-energy astrophysical
neutrinos have been considered in the past.  It has been noted that the
disappearance of all states except $\nu_1$ would prepare a beam that could in principle be used to measure elements of the neutrino mixing matrix, namely the ratios $U^2_{e1} : U^2_{\mu 1} : U^2_{\tau 1}$\cite{pakvasa2}.  
The possibility of measuring
neutrino lifetimes over long baselines was mentioned in Ref.\cite{weiler}, and some predictions for decay in four-neutrino models were given in Ref.\cite{keranen}.  We have shown that the particular values and small uncertainties on the neutrino mixing parameters allow for the first time very distinctive signatures of the effects of neutrino decay on the detected flavor ratios.  The expected increase in neutrino lifetime sensitivity (and corresponding anomalous 
neutrino couplings) by several orders of magnitude makes for a very
interesting test of physics beyond the Standard Model; a discovery would
mean physics much more exotic than neutrino mass and mixing alone.  As shown
below,  neutrino decay because of its unique signature cannot be mimicked by either different neutrino flavor ratios at the source or other non-standard neutrino interactions.

A characteristic feature of decay is its strong energy dependence: $exp (-Lm/E \tau)$, where $\tau$ is the rest-frame lifetime.  
For simplicity, we will assume that decays are always complete, i.e., that 
these exponential factors vanish.  The assumption of complete decay means we do not have to consider the distance and intensity distributions of sources.  We assume an isotropic diffuse flux of high-energy astrophysical neutrinos, and can thus neglect the angular deflection of daughter neutrinos from the trajectories of their parents\cite{lindner}.

{\bf Disappearance only}.  Consider the case of no detectable decay
products, that is, the neutrinos simply disappear.  This limit is interesting for decay to 'invisible'  daughters, such as a sterile neutrino, and also for decay
to active daughters if the source spectrum falls sufficiently steeply with
energy.  In the latter case, the flux of daughters of degraded energy will
make a negligible contribution to the total flux at a given energy.  Since coherence will be lost we have.
\begin{eqnarray}
\phi_{\nu_{\alpha}} = 
\sum_{i \beta} \phi^{source}_{\nu _\beta} (E) 
\mid U_{\beta i} \mid^2 \mid U_{\alpha i} \mid^2 e^{-L/\tau_i (E)} \\
 \stackrel{L \gg \tau_i}{\longrightarrow}  
\sum_{i(stable), \beta}  \phi^{source}_{\nu_{\beta}} (E) \mid U_{\beta i} \mid^2 \mid U_{\alpha i} 
\mid^2 ,
\end{eqnarray}
where the $\phi_{\nu_{\alpha}}$ are the fluxes of $\nu_\alpha, U_{\alpha i}$
are elements of the neutrino mixing matrix and $\tau$ are the neutrino
lifetimes in the laboratory frame.  Eq. (5) corresponds to the case where
decay is complete by the time the neutrinos reach us, so only the stable 
states are included in the sum.

The simplest case (and the most generic expectation) is a normal hierarchy in which both $\nu_3$ and $\nu_2$ decay, leaving only the lightest stable eigenstate  $\nu_1$.  In this case the flavor ratio is $U^2_{e1}:  
U^2_{\mu 1} : U^2_{\tau 1}$\cite{pakvasa2}. Thus if $U_{e3} = 0$
\begin{equation}
\phi_{\nu e} :  \phi_{\nu_{\mu}} : \phi_{\nu _{\tau}} = \cos^2 \theta_\odot :
\frac{1}{2}  \ \sin^2 \ \theta_\odot : \frac{1}{2} \sin^2 \theta_\odot \simeq
6 : 1 : 1, 
\end{equation}
where $\theta_\odot$ is the solar neutrino mixing angle, which we have set
to $30^0$.  Note that this is an extreme deviation of the flavor ratio from
that in the absence of decays.  It is difficult to imagine other mechanisms that would lead to such a high ratio of $\nu_e$ to $\nu_\mu$.  In the case of inverted hierarchy, $\nu_3$ is the lightest and hence stable state, and so
\begin{equation}
\phi_{\nu_{e}} :  \phi_{\nu_{\mu}} : \phi_{\nu _{\tau}} = U^2_{e3} : 
U^2_{\mu 3} : U^2_{\tau 3} = 0 : 1 : 1.
\end{equation}
If  $U_{e3} = 0$ and $\theta_{atm} = 45^0$, each mass eigenstate has equal
$\nu_\mu$ and $\nu_\tau$ components.  Therefore, decay cannot break the equality between the $\phi_{\nu_{\mu}}$ and $\phi_{\nu_{\tau}}$ fluxes and thus the $\phi_{\nu_{e}} : \phi_{\nu_\mu}$ ratio contains all the useful information. 
The effect of a non-zero $U_{e3}$ on the no-decay case of 1 : 1 : 1 is negligible.
For the normal hierarchy, the decay flavor mix  can change $\nu_e/\nu_\mu$
from 6/1 to 3/1 and $\nu_\mu/\nu_\tau$  from 1/1 to 3/1. In the case of
inverted hierarchy, there is no such dependence.

{\bf Appearance of daughter neutrinos}. --  If neutrino masses are quasi-degenerate, the daughter neutrino carries nearly the full energy of the parent.  An interesting and convenient feature of this case is that we can treat the effects of
the daughters without making any assumptions about the source spectra.
Including daughters of full energy, we have
\begin{eqnarray}
\phi_{\alpha}  (E) \stackrel{L \gg \tau_i}{\longrightarrow}
\sum_{i \beta} \phi^{source}_{\beta} (E) 
\mid U_{\beta i} \mid^2 \mid U_{\alpha i} \mid^2 \\
 + 
\sum_{ij \beta} \phi^{source}_{\beta} (E) 
\mid U_{\beta j} \mid^2 \mid U_{\alpha i} 
\mid^2  B_{j \rightarrow i} \nonumber
\end{eqnarray}
where $B$ is a branching fraction and stable and unstable states are denoted
henceforth by $i$ and $j$ respectively.

If instead the neutrino mass spectrum is hierarchical, the daughter neutrinos will be degraded in energy with respect to the parent, so that
\begin{eqnarray}
\phi_{\nu_{\alpha}} (E) \stackrel{L \gg \tau_i}{\longrightarrow}  
\sum_{i \beta} \phi^{source}_{\nu_\beta} (E) \mid U_{\beta i} \mid^2 
\mid U_{\alpha i} \mid^2 \\
+ \int_{E}^{\infty} \ d E' \ W_{E ' E} 
\sum_{i j \beta} \phi^{source}_{\nu_{\beta}} (E') \mid U_{\beta j} \mid^2 
\mid U_{\alpha i} \mid^2 B_{j \rightarrow i},  \nonumber
\end{eqnarray}
where $E$ is the daughter and $E'$ is the parent energy.  The normalized energy spectrum of the daughter is given by
\begin{equation}
W_{E' E} = \frac{1 } {\Gamma (E')}    
\frac{d \Gamma(E'. E)}{dE}
\end{equation}
If the neutrinos are Majorana particles, daughters of both helicities will be detectable (as neutrinos or antineutrinos), whereas if they are Dirac particles, daughters of one helicity will be sterile and hence undetectable.
In the hierarchical limit, the  energy distributions in the lab frame are
$E/E^{'2}$ and $(E'-E)/E^{'2}$ respectively for the helicity conserving and
helicity flipping modes.

In the case of Majorana neutrinos, we may drop the distinction between neutrino and antineutrino daughters and sum over helicities.  Assuming the source spectrum to be a simple power law, $E^{-\alpha}$, we find
\begin{eqnarray}
\phi_{\nu_{\alpha}}  (E) \stackrel{L \gg \tau_i}{\longrightarrow}  
\sum_{i \beta} \phi^{source}_{\nu_{\beta}} (E) \mid U_{\beta i} \mid^2 
\mid U_{\alpha i} \mid^2 \\
 + \frac{1}{\alpha} 
\sum_{i j \beta} \phi^{source}_{\nu_{\beta}} (E) 
\mid U_{\beta j} \mid^2 
\mid U_{\alpha i} \mid^2 B_{j \rightarrow i} \nonumber
\end{eqnarray}
This is identical to the expression in Eq.(8) except for the overall factor
of 1/$\alpha$ in front of the second term.  For Dirac neutrinos we detect
only the daughters that conserve helicity, the effect of which is only to
change the numerical coefficient of the second sum in Eq. (11). Thus, although the flavor ratio will differ from the cases above, it is still independent of energy-i.e., decay does not introduce a spectral distortion of the power law.  We stress that we have assumed a simple but reasonable power law spectrum $E^{- \alpha}$; a broken power law spectrum, e.g. would lead to a more complicated energy 
dependence.

{\bf Uniqueness of decay signatures.}--Depending on which of the mass eigenstates are unstable, the decay branching ratios, and the hierarchy of the neutrino mass eigenstates, quite different ratios result.  For the normal hierarchy, some possibilities are shown in Table~\ref{flavor}.

\begin{table}[h]
\begin{center}
\caption{Flavor ratios for various decay scenarios.}\label{flavor}
  \small
  \begin{tabular}{c|l|l|c}\hline\hline
\mbox{Unstable} & \mbox{Daughters} &  \mbox{Branching} & $\phi_{\nu_{e}}$ : 
$\phi_{\nu_{\mu}} : \phi_{\nu_{\tau}}$ \\ \hline\hline
$\nu_2, \nu_3$   & \mbox{anything}    & \mbox{irrelevant}   &6 : 1 : 1 
\\ \hline
$\nu_3$          & \mbox{sterile}     &  \mbox{irrelevant}  &2 : 1 : 1 
\\ \hline
$\nu_3$     & \mbox{full energy}  & $B_{3 \rightarrow 2} =1$  &1.4 : 1 : 1 
\\
     & \mbox{degraded} $(\alpha =2)$   &     
&1.6 : 1 : 1 \\ \hline 
$\nu_3$     & \mbox{full energy}  & $B_{3 \rightarrow 1} =1$  &2.8 : 1 : 1 
\\
     & \mbox{degraded} $(\alpha =2)$   &    
&2.4 : 1 : 1 \\ \hline
$\nu_3$     & \mbox{anything}    & $B_{3 \rightarrow 1}$ = 0.5   & 2 : 1 : 1 \\            &                    & $B_{3 \rightarrow 2}$ = 0.5     & \\ \hline
\hline
\end{tabular}
\end{center} 
\end{table}

The most natural possibility with unstable neutrinos is that the heaviest two mass eigenstates both completely decay.  The resulting flavor ratio is just that of the lightest mass eigenstate, independent of energy and whether daughters are detected or not.  For normal and inverted hierarchies we obtained 6 : 1 : 1 and 0 : 1 : 1 respectively.  Interestingly, both cases have extremes $\phi_{\nu_{e}} : \phi_{\nu_{\mu}}$ ratios, which provides a very useful diagnostic.  Assuming no new physics besides decay, a ratio greater that I suggests the normal hierarchy, while a ratio smaller than 1 suggests an inverted hierarchy.  In the case that decays are not complete these trends still hold. 

An important issue is how unique decay signatures would be.  Are there other scenarios (either non-standard astrophysics or neutrino properties) that would give similar ratios?  There exist astrophysical neutrino production models with different initial flavor ratios, such as 0 : 1 : 0 \cite{rachen}, 
for which the detected flavor ratios (in the absence of decay) would be about 0.5 : 1 : 1.  However, since the mixing angles $\theta_\odot$ and 
$\theta_{atm}$ are both large, and since the neutrinos are produced and
detected in flavor states, no initial flavor ratio can result in a measured
$\phi_{\nu_{e}} : \phi_{\nu_{\mu}}$ ratio anything like that of the two main cases, 6 : 1 : 1 and 0 : 1 : 1.

In terms of non-standard particle physics, decay is unique in the sense that
it is ``one way'', unlike, say, oscillations or magnetic moment transitions.  Since the initial flux ratio in the mass basis is 1 : 1 : 1, magnetic moment transitions between (Majorna) mass eigenstates cannot alter this ratio, due to the symmetry between $i \rightarrow j$ and $j \rightarrow i$ transitions.  On the other hand, if neutrinos have Dirac masses, magnetic moment transitions (both diagonal and off-diagonal) turn active neutrinos into sterile states, so the same 
symmetry is not present.  However, the process will not be complete in the
same way as decay--it will average out at 1/2, so there is no way would be left with a only single mass eigenstate.

{\bf Experimental Detectability.}--Deviations of the flavor ratios from 1 : 1 : 1 due to possible decays are so extreme that they should be readily identifiable\cite{beacom2}. Upcoming high energy neutrino experiments, such as Ice cube\cite{karte}, will not have perfect abilities to separarely measure the neutrino flux in each flavor.  However, the quantities we need are closely related to 
 observables, in particular in the limit of $\nu_\mu - \nu_\tau$ symmetry 
$(\theta_{atm} = 45^0$ and $U_{e3} = 0)$, in which all mass eigenstates contain equal fractions of $\nu_\mu$ and $\nu_\tau$.  In that limit, the fluxes for $\nu_\mu$ and $\nu_\tau$ are always in the ratio 1 : 1, with or without decay. 
This is useful since the $\nu_\tau$ flux is the hardest to measure. 

Detectors such as IceCube will be able to directly measure the $\nu_\mu$ flux
by long-ranging muons which leave tracks through the detector.  The
charged-current interactions of $\nu_e$ produce electromagnetic showers.
However, these are be hard to distinguish from hadronic showers caused by all flowers through their neutral-current  interactions, or from the charged-current interactions of $\nu_\tau$ (an initial hadronic shower followed by either an electromagnetic 
or hadronic shower from the tau lepton decay)\cite{ahrens}.  We thus consider our only experimental information to be the number of muon tracks and the number of showers. 

The relative number of shower events to track events can be related to the most interesting quantity for testing decay scenarios, i.e., the $\nu_e$ to 
$\nu_\mu$ ratio.  The precision of the upcoming experiments should be good enough to test such extreme flavor ratios produced by decays.  If electromagnetic 
and hadronic  showers can be separated, then the precision will be even better.

Comparing, for example, the standard flavor ratios of 1 : 1 : 1 to the
possible 6 : 1 : 1 generated by decay, the more numerous electron neutrino
flux will result in a substantial increase in the relative number of shower events.
The details of this observation depends on the range of muons generated in
or around the detector and the ratio of charged to neutral current cross
sections.  This measurement will be limited by the energy resolution of the
detector and the ability to reduce the atmospheric neutrino background.  The
atmospheric background drops rapidly with energy and should be negligibly small at and above the PeV scale.

{\bf Discussion and Conclusions.}-  The flux ratios we discuss are energy-independent because we have assumed that the ratios at production are energy-independent, that all oscillations are averaged out, and that all possible decays are complete.  The first two assumptions are rather generic, and the third is a reasonable simplifying assumption.  In the standard scenario with only oscillations, the final flux ratios are $\phi_{\nu_{e}} :  \phi_{\nu_{\mu}} : 
\phi_{\nu_{\tau}} = 1 : 1 : 1$.  In the cases with decay, we have found rather different possible flux ratios, for example 6 : 1 : 1 in the normal hierarchy and 
0 : 1 : 1 in the inverted hierarchy.  These deviations from 1 : 1 : 1 are so extreme that they should be readily measurable.

If we are very fortunate\cite{barenboin}, we may be able to observe a reasonable number of events from several sources (of known distance) and/or over a sufficient range in energy.  Then the resulting dependence of the flux ratio 
$(\nu_e/\nu_\mu)$ on L/E as it evolves from say 6 (or 0) to 1, can be clear
evidence of decay and further can pin down the actual lifetime instead of
just placing a bound.

To summarize, we suggest that if future measurements of the flavor mix at
earth of high energy astrophysical neutrinos find it to be
\begin{equation}
\phi_{\nu_{e}} / \phi_{\nu_{\mu}} / \phi_{\nu_{\tau}} = \alpha / 1 / 1 ;
\end{equation}
then
\begin{description}
\item[(i)] $\alpha =1$ (the most boring case) confirms our knowledge of the
MNS\cite{MNS} matrix and our prejudice about production mechanism;
\item[(ii)] $\alpha \approx 1/2$ indicates that the source emits pure
$\nu_\mu's$ and the mixing is conventional;
\item[(iii)] $\alpha > 1$ indicates that neutrinos are decaying with normal
hierarchy; and 
\item[(iv)]$\alpha \ll 1$ would mean that neutrino decays are occuring with inverted hierarchy.
\end{description}
  
\section{Acknowledgements}
  This talk is based on published and ongoing work in collaboration with
John Beacom, Nicole Bell, Dan Hooper, John Learned  and Tom Weiler. I thank them for a most
enjoyable collaboration.  I also thank Milla Baldo-Ceolin for
another memorable week in Venice.  This work was supported in part by U.S.D.O.E. under grant DE-FG03-94ER40833.

\end{document}